\documentclass[useAMS,usenatbib,usegraphicx]{mn2e}
 \usepackage[utf8]{inputenc}
 \usepackage[russian,english]{babel}
\title[Magnetic field and radial velocities of the star Chi Draconis A]
  {Magnetic field and radial velocities of the star Chi Draconis A}
\author[Byeong-Cheol~Lee et al.\foreignlanguage{russian}{}]
  {Byeong-Cheol~Lee$^{1,2}$,\thanks{bclee@kasi.re.kr}
  D. Gadelshin$^3$, Inwoo Han$^{1,2}$, Dong-Il Kang$^1$, Kang-Min Kim$^1$,
\newauthor
  G. Valyavin$^3$, G. Galazutdinov$^{4,3,5}$, Gwanghui Jeong$^{1,2}$, N. Beskrovnaya$^5$, T. Burlakova$^3$,
\newauthor
A. Grauzhanina$^3$, N.R. Ikhsanov$^{5,6}$, A.F. Kholtygin$^6$,
A. Valeev$^3$, V. Bychkov$^3$,
\newauthor
and Myeong-Gu Park$^7$\\
  $^1$Korea Astronomy and Space Science Institute 776, Daedeokdae-ro, Yuseong-gu, Daejeon 34055, Korea, bclee@kasi.re.kr\\
  $^2$Astronomy and Space Science Major, Korea University of Science and Technology, Gajeong-ro Yuseong-gu, Daejeon 34113, Korea\\
  $^3$Special Astrophysical Observatory, Russian Academy of Sciences,
  369167 Nizhnii Arkhyz, Russia\\
  $^4$Instituto de Astronomia, Universidad Catolica del Norte, Av. Angamos 0610,
  Antofagasta, 1270709 Chile\\
  $^5$Central Astronomical Observatory at Pulkovo, Russian Academy of Sciences, Pulkovskoe Shosse 65, 196140 Saint-Petersburg, Russia\\
  $^6$Saint-Petersburg State University, Universitetskij pr.\,28, 198504 St.\,Petersburg,  Russia\\
  $^7$Department of Astronomy and Atmospheric Sciences, Kyungpook National University, Daegu 41566, Korea
  }
\date{Released 2017 August 04}

\pagerange{\pageref{firstpage}--\pageref{lastpage}} \pubyear{2017}

\def\LaTeX{L\kern-.36em\raise.3ex\hbox{a}\kern-.15em
    T\kern-.1667em\lower.7ex\hbox{E}\kern-.125emX}

\begin{document}

\label{firstpage}

\maketitle

\begin{abstract}

We present high-resolution spectropolarimetric observations of the spectroscopic binary $\chi $\,Dra. Spectral lines in the spectrum of the main
component $\chi $\,Dra\,A show variable Zeeman displacement, which confirms earlier suggestions about the presence of a weak magnetic field on the surface
of this star. Within about 2 years of time base of our observations, the longitudinal component $B_{\rm L}$ of the magnetic field exhibits variation from $-11.5 \pm 2.5$\,G to $+11.1 \pm 2.1$\,G with a period of about 23 days. Considering the rotational velocity of $\chi $\,Dra\,A in the literature and that newly measured in this work, this variability  may be explained by the stellar rotation under the assumption that the magnetic field is globally stable.
Our new measurements of the radial velocities (RV) in high-resolution $I$-spectra of $\chi $\,Dra\,A refined the orbital parameters and reveal persistent deviations of RVs from the orbital curve. We suspect that these deviations may be due to the influence of local magnetically generated spots, pulsations, or a Jupiter-size planet orbiting the system.
\end{abstract}

\begin{keywords}
magnetic fields -- stars: individual: $\chi$ Dra: binaries.
\end{keywords}

\section{Introduction}

The spectroscopic binary system $\chi$\,Dra is a classic spectroscopic binary first discovered by \citet{Cam}. Since 1987 \citep{Tomkin,Schoeller} the
system is also known as an interferometric binary. The angular separation between components is $0\farcs12$, and the orbital period is 280.55 days
\citep{Tomkin,Schoeller}. The primary component $\chi $\,Dra\,A is a F7V 4th magnitude star with a projected rotational velocity of
$v \sin i = 2.5$\,km s$^{-1}$ \citep{Gray84b} \footnote{
\citet{Monin} lists erroneous value of $v \sin i = 11$\,km s$^{-1}$ in Table 1, citing \citep{Gray84a}.
}
and a radius of 1.2 $R_\odot$ \citep{Torres}. The secondary component is a convective K type star,  two  magnitudes fainter than the primary.

In a comparatively recent study by \citet{Monin}, it was suggested that the main component, $\chi $\,Dra\,A, has a weak longitudinal field of up to a
few tens of Gauss. This suggestion, along with the binarity of $\chi $\,Dra, makes this system an interesting laboratory to study the formation and
evolution of magnetic stars within multiple stellar systems. Motivated by this idea,  we  conducted an extensive set of high-resolution spectropolarimetric
observations of $\chi$\,Dra with spectropolarimetric facilities of the Bohyunsan Optical Astronomical Observatory (BOAO) of the Korea Astronomy and Space
Science Institute (KASI) in Republic of Korea. Another goal of this study was a high-precision search for the  radial velocity (RV) variations of the
system's main component. Observations, data reduction, and measurements are described in the next section. Section 3  presents results of magnetic field and
RV measurements. In Section 4, we discuss our findings.

\section{Observations, data reduction and measurements}

Observations of $\chi$\,Dra were carried out on 15 nights between 2006 and 2008. The BOES spectropolarimeter at the 1.8-m of the BOAO was used.
The spectrograph and spectropolarimetric observational procedures are described by \citet{Kangmin}. The instrument is a moderate-beam fiber-fed high-resolution spectrograph which incorporates 3 STU Polymicro fibers of 300, 200, and 80 $\mu$m core diameter
(corresponding spectral resolutions are $\lambda/\Delta \lambda $ = 30\,000, 45\,000, and 90\,000, respectively). We used a 3800 -- 10\,000\,$\rm\AA$ working wavelength range and a spectropolarimetric
mode provided with a spectral resolution  of 60\,000 by using two additional fiber-fed channels.
An overview of   observations is given in Table\,\ref{t1}, where we list the date of observations, total number of exposures,  typical exposure time for an individual frame, and  sky conditions.

\begin{table}
 \caption{ Observation log of the binary system $\chi$ Dra}
  \begin{tabular}{@{}cccc}
\hline
Date & N &  Exp(sec)& Sky conditions \\
\hline
27  Sep 2006   &                   4        &                360     &              good  \\
22  Jan 2007   &                   12       &                600     &              moderate \\
24  Jan 2007   &                   16       &                400     &              good \\
25  Jan 2007   &                   8        &                600     &              moderate \\
28  Jan 2007   &                   4        &                600     &              moderate  \\
31  Jan 2007   &                   6        &                480     &              good \\
02  Feb 2007   &                   8        &                500     &              moderate \\
03  Feb 2007   &                   24       &                360     &              good  \\
04  Feb 2007   &                    4       &                360     &              good  \\
25  Apr 2007   &                    4       &                360     &              good  \\
26  Apr 2007   &                    4       &                720     &              poor  \\
27  Apr 2007   &                    4       &                600     &              poor  \\
02  May 2007   &                   24       &                600     &              moderate \\
13  Jun 2008   &                    4       &                800     &              moderate \\
10  Sep 2008   &                    4       &                360     &              good  \\
\hline
\end{tabular}
\label{t1}
\end{table}

The reduction of spectropolarimetric data was carried out using the image processing program DECH
\footnote{ http://gazinur.com/DECH-software.html } as well as an automatic pipeline program \citep{Dongii1,Dongii2}. The general steps are standard and include cosmic ray hits removal, electronic bias and scatter light subtraction, extraction of the spectral orders, division by the flat-field spectrum, normalization to the continuum and wavelength calibration.

RV Measurements in the stellar spectra were carried out by calculating the
gravity centers of all narrow and symmetric spectral lines and their comparison with laboratory wavelengths which together with Lande factors and other atomic data necessary for data processing were derived from the VALD database \citep{Piskun,Ryabchik,Kupka}. Measurements of the longitudinal magnetic field were made as described in detail
by \citet{Kangmin}.

We identified several hundred lines in the spectrum of $\chi$\,Dra\,A within the range 3\,800 -- 10\,000\,$\rm\AA$ from which   we selected about 300
deepest ($r_c \ge 0.4$ where $r_c$ is the central depth of the line) single narrow and symmetric absorptions with non-zero Lande factors. By measuring
Zeeman displacements individually in all these lines, weighting and averaging these measurements as described by \citet{Monin}, we obtained the estimates
and corresponding uncertainties of the star's longitudinal magnetic field at paired exposures of different orientation of the quarter-wave plate following
the scheme described in detail by \citet{Kangmin}. Since the spin period of $\chi$\,Dra\,A is much longer than several days, we integrated these individual
estimates within a combined exposure for each observing nights. The duration of such exposure are equal to $N\times$Exp (see Table\,\ref{t1}) and ranges
from a few tens of minutes to a few hours.

In order to control the sign and zero level of the measured field, we used typical magnetic stars (HD~215441, HD~32633, and HD~40312) which have magnetic field of different strengths as well as zero-field stars (for details, see \citet{Kangmin}). To control our measurements, we used non-saturated telluric spectral lines and the star Procyon which has not demonstrated a magnetic field higher than one\,G \citep{Kangmin} and has a spectral class very similar to $\chi$\,Dra\,A .

Since individual RV estimates obtained for each short-time exposure have shown no  remarkable features, we averaged all   RV measurements   within combined
exposures in the same manner as  magnetic field estimates.

\begin{table}
 \caption{ Results of the radial velocity and magnetic field measurements for the star $\chi$ Dra\,A}
  \begin{tabular}{@{}cccccc}
\hline
Date & HJD-245 0000 & RV & $\sigma$ & $B_L$& $\sigma$ \\
  & (days)& (km s$^{-1}$) & (km s$^{-1}$) & (G) & (G)\\

\hline
27  Sep 2006 & 4006.096 &   38.78   &  0.05  &  +11.1   & 2.1   \\
22  Jan 2007 & 4123.337 &   36.351  &  0.017 &  +5.3    &  0.9  \\
24  Jan 2007 & 4125.334 &   34.836  &  0.013 &  -3.2    &  0.7   \\
25  Jan 2007 & 4126.331 &   34.112  &  0.021 &  -2.3    &  1.1   \\
28  Jan 2007 & 4129.320 &   32.126  &  0.029 &  -5.3    &  1.7   \\
31  Jan 2007 & 4132.345 &   29.898  &  0.032 &  -10.5   &  3.4  \\
02  Feb 2007 & 4134.093 &   27.019  &  0.017 &  -11.5   &  2.5   \\
03  Feb 2007 & 4135.345 &   26.089  &  0.018 &  -7.8    & 0.7   \\
04  Feb 2007 & 4136.347 &   24.421  &  0.045 &  -9.5    & 1.8    \\
25  Apr 2007 & 4216.014 &   25.58   &  0.046 &  +5.8    & 2.0    \\
26  Apr 2007 & 4216.999 &   25.55   &  0.045 &  +4.4    & 2.0   \\
27  Apr 2007 & 4218.061 &   26.31   &  0.045 &  +5.0    & 2.8    \\
02  May 2007 & 4223.218 &   26.88   &  0.023 &  -7.1    &   0.9  \\
13  Jun 2008 & 4630.817 &   46.56   &  0.035 &  -8.9    &  1.9   \\
10  Sep 2008 & 4719.988 &   12.24   &  0.040 &  -8.5    & 2.5    \\
\hline
\end{tabular}
\label{t2}
\end{table}

\section{Results}

Results of our measurements for each observing date are presented in Table\,\ref{t2} where column (1) is the date of observations, column (2) is a Heliocentric
Julian Date of mid-exposure, columns (3) and (4) display a nightly mean value of RV and its uncertainty $\sigma$, columns (5) and (6) give corresponding estimates of the longitudinal magnetic field $B_L$ and its uncertainty $\sigma$. Let us consider the results of Zeeman and RV measurements independently.

\subsection{Magnetic field}

\begin{figure}
\hspace{-0.cm}
\includegraphics[width=80mm,angle=0]{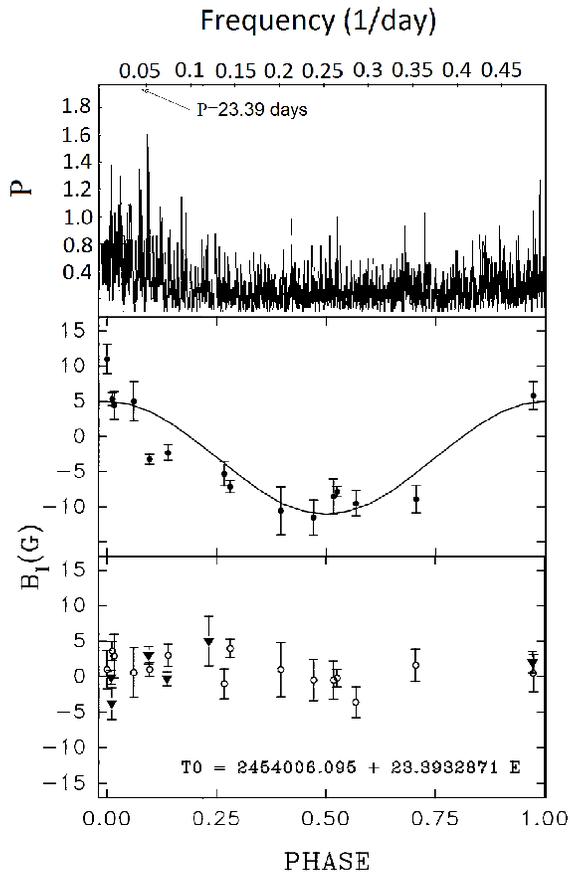}
\vspace{-0cm}
\caption{{\it Upper panel}: power spectrum of the field variation. {\it Middle panel}: observed phase variation of the longitudinal field $B_L$ of the
star $\chi$ Dra\,A (filled circles) and the best sinusoidal fit to the data (a solid line).  {\it Lower panel}: control results of ``zero-field''
measurements in the spectra of  Procyon (filled triangles) and telluric lines in the spectra of $\chi$ Dra\,A (open circles)}
\label{f1}
\end{figure}

As clearly seen from Table\,\ref{t2}, ten out of fifteen estimates of the longitudinal field have a significance exceeding the 3$\sigma$ level. Besides,
a long observing run between 22 Jan. 2007 and 04 Feb. 2007 revealed  monotonous change of the field $B_L$ from maximum to minimum
with passing through the negative extremum. This may indicate the presence of a global magnetic field on the stellar surface. Due to the rotation of
$\chi$ Dra\,A, the magnetosphere demonstrates different integral projections of the surface
magnetic field to the line of sight. If the magnetosphere is stable, this process should be periodic with the spin period of the star.
The two-year time base of our observations makes it possible to primarily investigate this possibility.

In order to determine the time-scale over which the longitudinal magnetic field varies (which may be the rotation period of the star),
we applied the Lafler-Kinman method. The method tests trial periods by requiring the sum of the squares of the field magnitude differences between
observations of adjacent phase to be a minimum. Such a criterion effectively reveals smooth periodical signals on a limited number of observations.
Analysis of the power spectrum (Fig.\,\ref{f1}, upper plot) of the data obtained with the Lafler-Kinman revealed a strong signal indicating the
presence of a period P = 23.39(9)\,days.

The magnetic phase curve of $\chi$ Dra\,A constructed for the found period is shown in Fig.\,\ref{f1} (middle plot). The phase variation of the
longitudinal field is symmetric with some deviations from the sinusoidal symmetry (for example two points
between $\phi = 0$ and $\phi = 0.25$). If from minimum to maximum individual measurements of Bl vary from $-11.5 \pm 2.5$\,G to $+11.1 \pm 2.1$\,G,
the sinusoidal fit of these data by the Marquard $\chi^2$ minimization method \citep{Bevington} gives the field
variation from $-11.5\,\pm 1.5$\,G to $+5.2\,\pm 1.5$\,G. This discrepancy suggests, that the field geometry is more complicated than
a simple dipole. The moment of maximum of the mean longitudinal field can be calculated according to the following ephemeris:
$T0 = 2454006.095 + 23.39(9)\,E$.

\subsection{Radial velocities}

Good quality of the RV data owing to the high mechanical stability of the BOES makes it   possible to reanalyse RVs of $\chi$ Dra\,A with an accuracy higher than achieved in previous studies. To the best of our knowledge, the most complete set of RV data for the system was presented and analysed by \citet{Tomkin}.
Using several tens of individual RV measurements as well as speckle observations, these authors determined the RV orbit for $\chi$ Dra. In more recent studies   \citep{Schoeller,Farrington}, the orbit was further refined by interferometric method. Combining RV measurements published by \citet{Tomkin} with our new measurements described here, we get a modified orbital  solution. The best-fit results are summarized in Table\,\ref{t3}, which lists
the projected velocity semi-amplitude of $\chi$ Dra\,A ($K$), the periastron angle ($W$), the epoch of periastron ({\bf $T_p$}), the orbital period ($P$),
the eccentricity ($e$), the offset RV ($V0$), and the linear slope $S$ of RV variation to remove the linear component of the variation.

\begin{table}
 \caption{ Radial-velocity orbit of $\chi$ Dra\,A}
  \begin{tabular}{@{}cccc}
\hline
Orbital element & Unit & Value & Uncertainty \\
\hline
$K$ & km s$^{-1}$ & 16.95 &   0.19  \\
$W$ &  deg($^{\circ}$) & 116.9 &   2.0 \\
$T_p$ & HJD & 2437307.827 &   1.294 \\
$P$ & days & 280.523 &   0.019 \\
$e$ &    &  0.432 &   0.02 \\
$V0$ & km s$^{-1}$ & 28.73 &  11.18 \\
$S$ & m s$^{-1}$day$^{-1}$ & 0.13 & 0.05 \\
\hline
\end{tabular}
\label{t3}
\end{table}

\begin{figure}
\hspace{-0.5cm}
\includegraphics[width=80mm,angle=0]{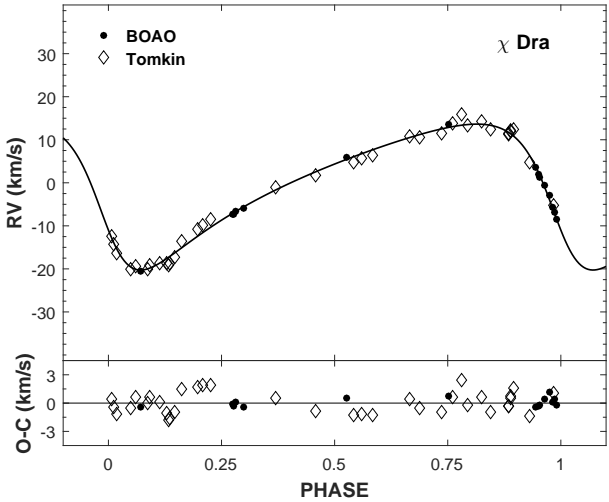}
\caption{{\it Upper panel}: observed radial velocities and constructed orbital curve. Open diamonds are data from \citet{Tomkin}. Filled circles
represent our data. {\it Lower panel}: residual radial velocities (the ``O-C'' values) after subtraction of orbital curve from the individual RV estimates.
}
\label{f2}
\end{figure}

Figure.\,\ref{f2} (upper plot) illustrates our solution (solid line) with all known RV estimates for $\chi$ Dra\,A folded with the obtained value of the orbital period P = 280.523\,days.
Filled circles represent our measurements and open diamonds data from \citet{Tomkin}. The standard deviation of the ``O-C'' (Observed minus Calculated) values (lower plot in the Fig.\,\ref{f2})
for the whole data set is 0.97 km s$^{-1}$, and for our measurements it is 0.37 km s$^{-1}$. This means that our data demonstrate much better agreement with the orbital curve thanks to the essential improvement in spectroscopic techniques since the 1980s. However, despite improved agreement, both our data and the data from \citet{Tomkin} demonstrate noticeable deviations of measured RVs from the calculated RV orbit. Considering that the orbital elements derived  in this paper (Table\,\ref{t2}) are in excellent agreement with the latest speckle-interferometric data \citep{Farrington},
we suspect that these deviations are not due to uncertainties in the orbital solution. For example, inspecting our RV estimates obtained within the  longest two-week observing run (22 Jan. -- 4 Feb. 2007) reveal an unexpectedly high scatter of the data (several times larger than observational uncertainties).
Visual inspection of Fig.\,\ref{f2} hints a periodicity within days to tens of days in the deviations of observed RVs from the orbital curve.
A similar picture can be seen in Fig.\,\ref{f2} from  \citet{Tomkin}. This suggests the presence of  additional cause of RV variations in $\chi$ Dra\,A. In order to examine this variability, we have analyzed the  ``O-C'' residuals from the spectroscopic orbital solution obtained here for all   available RV measurements.

\begin{figure}
\hspace{-0.4cm}
\includegraphics[width=85mm,angle=0]{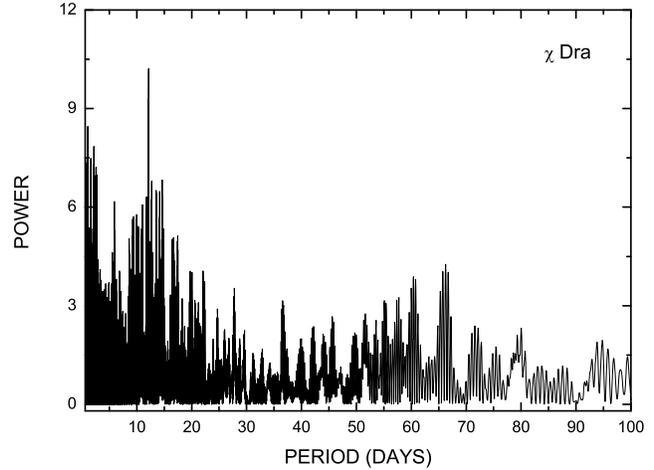}
\vspace{-0cm}
\caption{Power spectrum of the residual of radial velocity variations of $\chi$ Dra\,A. }
\label{f3}
\end{figure}

The Lomb-Scargle power spectrum \citep{Lomb, Scargle} of these ``O-C'' residuals is presented in Fig.\,\ref{f3}.
A considerable peak at $\sim$ 12 days was found to suggest the presence of a periodical signal.
Unfortunately, due to the strong inhomogeneity of the data, the very long time base of observations (tens of years), and insufficient amount of data limit ourselves to the illustration of the periodogogram only; we are presently unable to to clearly identify the true periodicity in the residual RVs of $\chi$ Dra\,A.
Additional high-precision spectral observations are needed for more reliable conclusions.

\section{Discussion}

We  obtained high-resolution spectropolarimetric observations of the star $\chi$ Dra\,A. Analysis of these new and previously published data
revealed the presence of variable longitudinal magnetic field. Within the 2-years time base of our observations the field varied from
$-11.5\,\pm 1.5$\,G to $+5.2\,\pm 1.5$\,G with the period P = 23.39(9)\,days.

As discussed by \citet{Tomkin} and \citet{Torres}, the star $\chi$ Dra\,A is a low mass, low metallicity old star. As such, the origin of the field on
$\chi$ Dra\,A should be typical for low mass stars of spectral classes from late F to cooler classes \citep{Reiners}. For these stars, as it is for
the Sun, magnetic fields are concentrated mainly into locally generated, dynamically unstable strong-magnetic tubes seen as dark spots on stellar surfaces.
These spots monotonously migrate with different velocities, giving additional contribution to the field variation in addition to rotation.
The found period P = 23.39\,days is in principle consistent with typical rotation periods of low mass stars
with masses comparable to $\chi$ Dra \,A, although it may be a bit longer than expected based on our measured longitudinal magnetic field strengths \citep{Marsden}. From this point of view it is important
to establish whether the star's rotation period is indeed around 23 days.

In order to clarify the situation with rotation we have analysed Doppler widths of spectral lines in the spectrum of $\chi$ Dra \,A. To measure the
projected rotational velocity $v \sin i$, we have chosen several single lines with small Lande factors. By modeling profiles of these lines using the
ATLAS/WIDTHS atmosphere model programs \citep{Kurucz}, we derived $v \sin i \le $\,3\,km/sec, which is consistent with
the estimate $v \sin i=\,2.5$\,km/sec by \citet{Gray84b}.  Surprisingly, $v \sin i=\,2.5$\,km/sec with the orbital inclination of about 75$^\circ$ \citep{Tomkin} and the stellar radius of 1.2 $R_\odot$ \citep{Torres} yields the rotation period of 23.5 days, almost the same as the
found period of P\,=\,23.39 days.

Thus, we suspect that the found 23.39 days period is mainly due to the rotation of the star. This result,
if confirmed, may also imply the existence of a long-living (more than several years)
global poloidal magnetic field. In contrast to the solar-type stars' unstable magnetic fields, stable poloidal (say dipolar) morphology
of the field suggests that we may be seeing a special case of fossil or generated magnetic field, originated and evolving within the frame of
the binary system. However, this interesting possibility is based on our currently limited observational data, and we don't exclude the
possibility that this variation could have more complicated origin and may not be regular. The detailed interpretation of the nature of the magnetic
field in $\chi$ Dra \,A requires further accumulation of observational data on longer time base. In this paper we restrict ourself with presentation of
new observational data confirming the presence of magnetized field structures on the surface of $\chi$ Dra\,A.

 Lastly, measured RVs of $\chi$ Dra\,A exhibit systematic deviations from the orbital curve.
Despite the fact that the measurements presented in this paper demonstrate improved agreement with the orbital solution, the  deviation still exists. No explanation of this phenomenon has been found so far. It may result from additional line displacement
due to magnetic nature of the star. For example, the presence of magnetic field with  inhomogeneous distribution over the stellar surface (in particular, magnetically-induced spots) may simply distort integral symmetry of spectral lines. Rotational modulation of such line profiles
can, in turn,  cause ``artificial'' RV variations.
However, the  period found through the analysis of RV residuals is not consistent with the rotation period estimated by means of  spectropolarimetric methods. Our data cannot exclude the existence of a hot Jupiter mass orbiting $\chi$ Dra\,A with a short period. Following this idea and taking into account that the
system is seen almost  edge-on, it seems reasonable to  monitor $\chi$ Dra\,A photometrically in order to search for deep, 1 -- 2\%, transit. New high-precision, high-resolution spectral observations of the star $\chi$ Dra\,A are also necessary  to answer  this particularly important question.

\section{Acknowledgements}

We thank the anonymous referee for useful comments. The authors acknowledge the Russian Science Foundation (grant N14-50-00043) for financial support of
theoretical part of this study. BCL acknowledges partial support by Korea Astronomy and Space Science Institute (KASI) grant 2017-1-830-03.
NGB acknowledges the support of the RAS Presidium Program P-7. AFK thanks the RFBR grant 16-02-00604~A for support of experimental investigaition of
the observed data. Support for MGP was provided by the KASI under the R\&D program supervised by the
Ministry of Science, ICT and Future Planning and by the National Research Foundation of Korea to the Center for Galaxy Evolution Research
(No. 2012-0027910).

\end{document}